\documentclass[fleqn,twoside]{article}

\usepackage[headings]{myespcrc2}

\usepackage{graphicx,epsfig, psfrag}
\usepackage[figuresright]{rotating}

\usepackage{amsmath}
\newcommand{\lfrac}[2]{\textstyle\frac{#1}{#2}\displaystyle}

\newcommand{\ntrl}[1]{\tilde{\chi}^0_#1}
\newcommand{\chrg}[1]{\tilde{\chi}^+_#1}

\newcommand{\snu}{\tilde{\nu}_e}

\def\mwrap#1{#1}
\def\dZneuL{\mwrap{\delta Z_{\tilde\chi^0}}}
\def\dZfneuLl(#1#2){\mwrap{\left[\delta Z_{\tilde\chi^0}\right]_{#1#2}}}
\def\dZfneuLlC(#1#2){\mwrap{\left[\delta Z_{\tilde\chi^0}\right]^\ast_{#1#2}}}
\def\dZfneuRl(#1#2){\mwrap{\left[\delta Z^{\rm R}_{\tilde\chi^0}\right]_{#1#2}}}
\def\dZchaL{\mwrap{\delta Z^{\rm L}_{\tilde\chi^+}}}
\def\dZchaR{\mwrap{\delta Z^{\rm R}_{\tilde\chi^+}}}
\def\dZfchaLl(#1#2){\mwrap{\left[\delta Z^{\rm L}_{\tilde\chi^+}\right]_{#1#2}}}
\def\dZfchaRl(#1#2){\mwrap{\left[\delta Z^{\rm R}_{\tilde\chi^+}\right]_{#1#2}}}

\def\dZSfl(#1#2){\mwrap{\left[\delta Z^{\tilde f}\right]_{#1#2}}}
\def\dZSflNI{\mwrap{\delta Z^{\tilde f}}}

\def\dZZAl(#1#2){\mwrap{\left[\delta Z_{\rm ZA}\right]_{#1#2}}}

\title{One-loop calculations for SUSY processes} 

\author{T. Fritzsche and W. Hollik
      \address{Max-Planck-Institut f\"ur Physik (Werner-Heisenberg-Institut)
       \hfill {\bf MPP-2004-83}   \\
       F\"ohringer Ring 6, D 80805 Munich, Germany}}
        
\begin{document}

\begin{abstract}

Strategy and results for complete one-loop computations 
in the Minimal Supersymmetric Standard Model are reviewed,
with applications to the calculation of SUSY mass spectra
and SUSY-particle processes. Determination of renormalization
constants and counterterms  are described in the on-shell 
renormalization scheme, and a translation between 
$\overline{\rm DR}$ and on-shell parameters is given.
As an example, 
cross sections for chargino and neutralino 
pair production in $e^+e^-$ annihilation are presented,
complete at the one-loop level.
\vspace{1pc}
\end{abstract}

\maketitle

\section{Introduction}

Experiments at future high-energy colliders will be able to discover
supersymmetric particles and to investigate their properties. 
A linear electron-positron
collider will be the best environment for precision studies
of supersymmetric models~\cite{TESLA}, especially of the minimal
supersymmetric standard model (MSSM). 
From precise measurements of masses and cross sections,
the fundamental parameters of the
MSSM Lagrangian can be reconstructed~\cite{reconstruct} to shed light 
on the mechanism of SUSY breaking. 
Adequate theo\-retical
predictions matching the experimental accuracy require the proper 
inclusion of higher-order terms in the calculations of mass spectra,
cross sections, decay rates 

In this talk we present a brief overview on complete
one-loop calculations for SUSY processes and supersymmetric
mass spectra,
where all MSSM particles 
with electroweak couplings are included in the
virtual states. As a calculational frame, the on-shell 
renormalization scheme has been chosen where all particle
masses are defined as pole masses, i.e.\ on-shell quantities.  
Cross sections are thus directly related to the physical masses of
the external particles  and of the other particles entering the loops.  
We outline the theo\-retical basis of the calculation and 
show in a few examples the numerical size of the loop effects, where
we restrict ourselves to the case of the 
CP-conserving MSSM with real parameters.
The presentation is based on~\cite{Fritzsche:2002bi,Hollik:2003jj}, 
for other approaches see~\cite{others1,others2}.

\section{Renormalization}

\subsection{Charginos and neutralinos}

The bilinear part of the Lagrangian describing the 
chargino/neutralino sector of the MSSM
involves the $\mu$ parameter,  
the soft-breaking gaugino-mass parameters $M_1$ and $M_2$, 
and the Higgs vacua $v_i$, which are related to 
$\tan\beta = v_2/v_1$ and to the $W$ mass
$M_W = g v/2$ with $v = (v_1^2 + v_2^2)^{1/2}$.

Renormalization constants are introduced for the chargino mass matrix $X$
and for the chargino fields $\chrg{i}$ $(i=1,2)$ by the transformation
\begin{eqnarray}
X & \to & X + \delta X \, , \nonumber \\
\omega_L\,\chrg{i} & \to & 
\left(\delta_{ij}+\lfrac{1}{2}\dZfchaLl(ij)\right)\,\omega_L\,\chrg{j}\, ,
\nonumber \\
\omega_R\,\chrg{i} & \to & 
\left(\delta_{ij}+\lfrac{1}{2}\dZfchaRl(ij)^\ast\right)\,\omega_R\,\chrg{j}
\, .
\label{eqn:RvRChar2}
\end{eqnarray}
The matrix $\delta X$ consists of the counterterms for the 
parameters in the mass matrix $X$,
\begin{eqnarray}
\delta X  = \!  \left(\!\!\!  \begin{array}{cc}
\delta M_2 & \sqrt{2}\,\delta\big(M_W\,\sin\beta\big)\\
\sqrt{2}\,\delta\big(M_W\,\cos\beta\big) & \delta\mu
\end{array}\!\!\! \right)\! .  \hspace{-5cm}
\label{eqn:RenChar1aa}
\end{eqnarray}
The field-renormalization constants $\dZchaL$ and $\dZchaR$ 
are general complex 2$\times$2 matrices.
The set of renormalization constants in~(\ref{eqn:RvRChar2})
renders both 
$S$-matrix elements and Green functions for charginos finite
and allows moreover to get the matrix of renormalized self energies
diagonal on the mass shell of each of the chargino mass eigenstates 
$\tilde{\chi}^+_{1,2}$.

The neutralino fields $\ntrl{i}$ $(i=1,\ldots,4)$ and the
mass-matrix $Y$ are -- in analogy to the chargino case --
renormalized by the substitutions
\begin{eqnarray}
Y & \to & Y + \delta Y \, , \nonumber\\
\omega_L\,\ntrl{i} & \to & 
\left(\delta_{ij}+\lfrac{1}{2}\dZfneuLl(ij)\right)\,\omega_L\,\ntrl{j}
 \, ,\nonumber \\
\omega_R\,\ntrl{i} & \to & 
\left(\delta_{ij}+\lfrac{1}{2}\dZfneuLlC(ij)\right)\,\omega_R\,\ntrl{j}
\;\;.
\label{eqn:RvRNeutr2b}\label{eqn:RvRNeutr2}
\end{eqnarray}
Besides those parameter counterterms already present in
(\ref{eqn:RenChar1aa}),
the counterterm matrix $\delta Y$ contains
the counterterms for 
the soft-breaking gaugino-mass parameter $M_1$,
for the $Z$ mass and the electroweak mixing angle,
respectively.
The matrix-valued renormalization constant
$\dZneuL$ is a general complex 4$\times$4 matrix. 

Using the on-shell approach of \cite{Fritzsche:2002bi},
the pole masses of the two charginos, $m_{\chrg{1}}, m_{\chrg{2}}$, and
of one neutralino, $m_{\ntrl{1}}$, are considered as input parameters,
to specify the chargino/neutralino Lagrangian in terms of physical quantities. 
This is equivalent both to the specification of the parameters $\mu, M_1,
M_2$, which are related to the input masses in the same way as in lowest
order, as a consequence of the on-shell renormalization conditions,
and to the definition of the respective counterterms.
In that way, the tree-level masses $m_{\chrg{i}}$ and $m_{\ntrl{j}}$
as well as the counterterm matrices
$\delta X$ and $\delta Y$ are fixed.

Furthermore, one requires for both, charginos as well as neutralinos,
that the matrix of the renormalized one-particle-irreducible two-point
vertex functions $\hat{\Gamma}^{(2)}_{ij}$ becomes diagonal for on-shell
external momenta. This fixes the non-diagonal entries of the
field-renormalization matrices $\dZchaL$, $\dZchaR$, and $\dZneuL$; 
their diagonal entries are determined by normalizing the residues of the
propagators to be unity.  

\subsection{Sfermions}

In the sfermion sector, we introduce renormalization constants for the
mass matrices  in the $L,R$ basis for each sfermion species,
$M^{\tilde f}$, 
and for the mass-eigenstates 
${\tilde f}^s$ ($s=1,2$ in general, $s=L$ for $f=\nu$) 
by means of the transformation 
\begin{eqnarray}
M^{\tilde f} & \to & M^{\tilde f} + \delta M^{\tilde f} \, ,\nonumber \\
{\tilde f}^s & \to & 
\left(\delta_{st}+\lfrac{1}{2}\dZSfl(st)\right)\,{\tilde f}^t
\;\;.
\label{eqn:RenSelectron1}
\end{eqnarray}
The matrix $\delta M^{\tilde f}$ is made of the counterterms for the 
parameters in the mass matrix $M^{\tilde f}$.
In the case of sneutrino fields, the respective mass matrix actually
consists of only one single element. 
The field-renormalization constants 
$\dZSflNI$ are general complex 2$\times$2 matrices, except for the
$f=\nu$ case with a single field-renormalization constant 
$\delta Z^{\tilde{\nu}} =\delta Z_L$ only.
 
Applying the renormalization procedure of \cite{Hollik:2003jj}, 
the sfermion soft-breaking $L,R$ mass parameters for the 
members of each isodoublet, together with their renormalization 
constants, are determined via on-shell mass renormalization
conditions for three independent pole masses in the case of squarks
and two in the case of sleptons.
The trilinear couplings $A_f$ can be related to the non-diagonal
entries of the field-renormalization matrices $\dZSflNI$,
and are determined via the requirement of having diagonal
two-point vertex functions for on-shell momenta, as in the case 
of charginos and neutralinos.
The diagonal entries of $\dZSflNI$ are determined by
normalizing the residues of the propagators to unity.  

\subsection{Other sectors}

The formal description of parameter and field renormalization
in the Standard-Model-like part of the MSSM is taken over
from~\cite{Denner:kt}, yielding for 
the electric charge 
\begin{eqnarray}
e & \to & Z_e\,e = (1+\delta Z_e)\,e \, ,
\label{eqn:RenOther1}
\end{eqnarray}
and for the 
masses of $W$, $Z$, and of the fermions 
\begin{eqnarray}
M_{W,Z}^2 & \to & M_{W,Z}^2 + \delta M_{W,Z}^2 \, , \nonumber \\
m_{f}& \to & m_{f} + \delta m_f \, .
\label{eqn:RenOther2}
\end{eqnarray}
The counterterms are determined by the usual on-shell conditions
in each case;
field renormalization is treated as 
described in~\cite{Denner:kt}.

Renormalization of $\tan\beta$ as the ratio  
of the VEVs of the two Higgs fields is done in the 
$\overline{\rm DR}$-scheme~\cite{tanbetarenormalization},
\begin{eqnarray}
\tan\beta = \frac{v_2}{v_1}  & \to &
\frac{v_2}{v_1} \left( 1+\frac{1}{2} 
                (\delta Z_{H_2}-\delta Z_{H_1}) \right)
\label{eqn:RenOther4}
\end{eqnarray}
with the $\overline{\rm DR}$-singular parts of   
the Higgs-field renormalization constants $\delta Z_{H_{1,2}}$.
 
\section{$\overline{\rm DR}$ versus $\rm OS$ scheme}
\label{sec:DR2OS}

In the $\overline{\rm DR}$ scheme, divergent 1-loop
quantities are renormalized by adding counterterms 
that are proportional to the divergent parts,
$$ \frac{2}{\epsilon}-\gamma+\log 4\pi \, , $$
of the 2- and 3-point vertex functions,
regularized using the dimensional-reduction method.
As a consequence, physical observables depend 
on the scale $\mu_{\overline{\rm DR}}$.
Input variables in the $\overline{\rm DR}$ scheme are
a natural choice for GUT-inspired parameter sets (e.g.~in SUGRA
scenarios).

On the other hand, in the ${\rm OS}$ scheme the renormalization
constants are fixed at physical scales; observables are thus scale
independent. The ${\rm OS}$ scheme is convenient for
calculations of cross sections and decay rates, because 
masses at Born level and in higher order agree (with few exceptions),
holding the correct phase-space kinematics already in
tree-level calculations. 

In order to achieve a translation
between $\overline{\rm DR}$ and OS parameters, the following two steps 
are performed, specified here for the quantities
$\mu$, $M_2$, $M_1$ 
of the chargino/neutralino sector.  
\begin{enumerate}
\item Using  $\mu$, $M_2$, and $M_1$ in the $\overline{\rm DR}$ scheme as
  a starting point,
  the pole masses of three particles, e.g.\  
  both charginos and the
  lightest neutralino, are calculated at the one-loop level.
\item From those three physical masses the corresponding parameters in
  the ${\rm OS}$ scheme are deduced, using tree-level relations (which
  are left unaltered by construction in the ${\rm OS}$ scheme).
\end{enumerate}

\noindent
In the example below we show the values (in GeV) for the various quantities
using the SPS1a set~\cite{sps1a}
of input parameters in the $\overline{\rm DR}$ 
scheme, where shifts up to 10 GeV can occur.\\

\noindent
\renewcommand{\tabcolsep}{0.5pc} 
\renewcommand{\arraystretch}{1.2} 
\begin{tabular}{@{}rrr}
\hline
$\overline{\rm DR}$\hspace{0.6cm} & pole masses \hspace{0.1cm}& 
${\rm OS}$\hspace*{0.6cm}\\
\hline
$\mu=352.4$ &            $m_{\chrg{1}}=184.3$ & $\mu=361.6$\\
$M_1=\hphantom{0}99.0$ & $m_{\chrg{2}}=387.4$ & $M_1=102.4$\\
$M_2=192.7$ &            $m_{\ntrl{1}}=\hphantom{0}99.5$ & $M_2=201.0$\\
\hline
\end{tabular}

\section{\boldmath{$e^+e^-$} production cross sections}

As a concrete example, 
we discuss the production of chargino and neutralino pairs 
in $e^+e^-$ annihilation (see also \cite{pairproduction}).
For production of sfermion pairs we refer to the 
literature~\cite{sfermionpairs}.

\subsection{Born amplitudes}

At lowest order, 
the amplitude $\mathcal{M}$ for chargino-pair production can be
described by $s$-channel photon and $Z$-boson exchange and by
$t$-channel exchange of a scalar neutrino $\snu$, as displayed in the  
following  Feynman diagrams.
\begin{center}
\vspace{0.2cm}
\epsfig{file=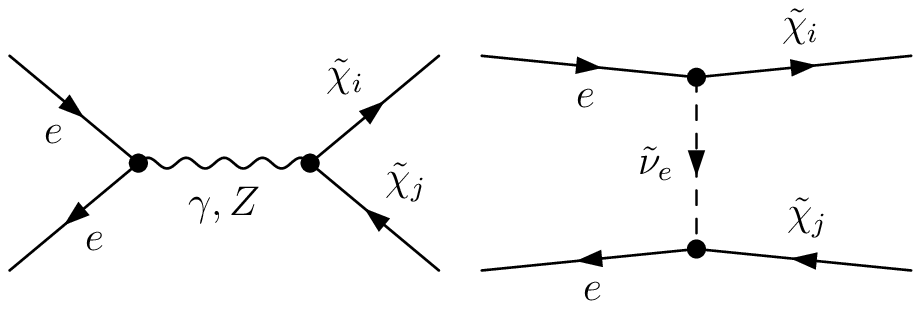,width=0.8\linewidth}
\\[0.2cm]
\end{center}
In the case of neutralino pair production, there is no
photon exchange at tree level, and $t$-channel exchange 
is mediated by the two selectrons $\tilde e^s$, $s=1,2$.
\begin{center}
\vspace{0.2cm}
\epsfig{file=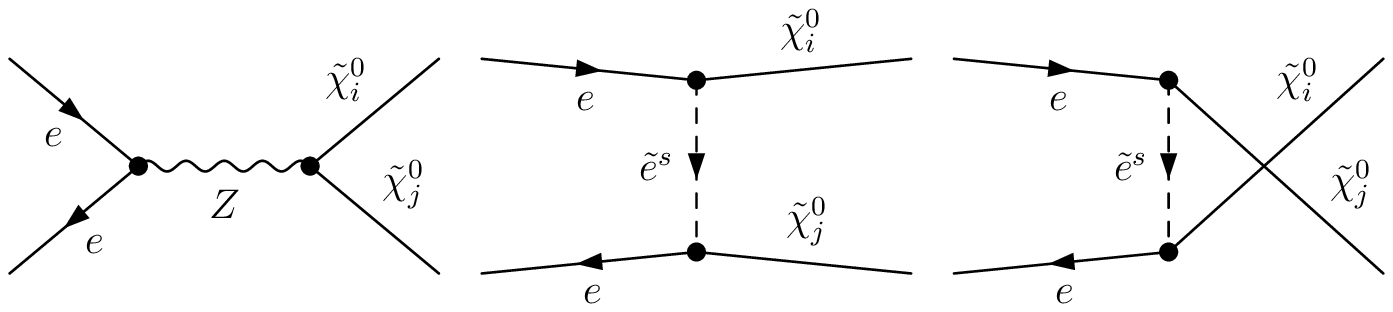,width=0.99\linewidth}
\\[0.2cm]
\end{center}
Other diagrams, containing Higgs lines, are always negligible.

\subsection{Virtual corrections}
The above set of Born diagrams has to be dressed by the corresponding
loop contributions containing the full particle spectrum of the MSSM.
The renormalization constants determined in section 2 
are complete to deliver all
counterterms required for propagators and vertices appearing in
the amplitudes, and they have become part of the {\tt FeynArts}
package for the MSSM~\cite{feynarts}.
The one-loop contributions can be classified as follows,
\begin{itemize}
\item {\sf self-energy contributions} to the propagators 
  for photon, $Z$ boson, and 
   $\tilde\nu_e$ ($\tilde e^s$) sneutrino (selectrons);
\item {\sf vertex corrections} for $\gamma$ and $Z$ in the
  $s$ channel and $\tilde\nu_e$ ($\tilde e^s$) in the $t$ channel, 
\item {\sf box-diagram contributions} with double exchange of gauge, Higgs,
    and SUSY particles in the $s$ and $t$ channel.
\end{itemize}

\subsection{Real photons and ``QED corrections''}
\label{subsec:Bremsstrahlung}

Virtual photons 
attached to external charged particles
give rise to infrared (IR) divergences in the loop diagrams.
An IR-finite result is obtained by adding real-photon bremsstrahlung
integrated over the photon phase space.
The sum of the one-loop contribution to the cross section,
$\sigma^{\rm virt}$, and the bremsstrahlung cross section, 
$\sigma^{\rm brems}$, is IR-finite.
Different from standard-fermion production, these photonic contributions
cannot be removed and treated separately as ``QED corrections'',
because the one-loop result without the virtual photons would not
be UV-finite.
The presence of the photon is required to cancel the divergence from
the photino component of the virtual neutralinos. 

For cancellation of the IR divergence, it is convenient to split
$\sigma^{\rm brems}$ into a (IR-divergent) soft part 
and a (IR-finite) hard part, both  depending on 
a  soft-photon cutoff $\Delta E$ for the energy of the radiated photon,
\begin{eqnarray}
\sigma^{\rm brems} & = & 
\sigma^{\rm soft}(\Delta E) + \sigma^{\rm hard}(\Delta E)
\;\;.
\end{eqnarray}
For practical calculations, in order to avoid numerical instabilities,
$\sigma^{\rm hard}$ is divided 
into a collinear part,
where the photon is within an angle smaller than 
$\Delta\theta$ with respect to the radiating particles,
and the complementary non-collinear part,
\begin{eqnarray}
\sigma^{\rm hard} & = & 
\sigma^{\rm coll}(\Delta \theta) 
+ \sigma^{\rm non-coll}(\Delta \theta) \, .
\end{eqnarray}
For $\sigma^{\rm soft}$ and $\sigma^{\rm coll}$ 
analytical results are available;
$\sigma^{\rm non-coll}$ is calculated numerically
with the help of {\tt DIVONNE} and {\tt CUHRE},
multidimensional numerical integration
routines that are part of the CUBA-library \cite{Hahn:2004fe}.

As already pointed out above,
there is no diagrammatic way to disentangle QED-like photonic 
virtual contributions from the MSSM-specific parts.
One can, however, isolate the universal and leading
QED terms in $\sigma^{\rm virt} + \sigma^{\rm soft}$
resulting from photons collinear to the incoming $e^{\pm}$,
which contain the large logarithm
$L_e =\log \left(\frac{s}{m_e^2}\right)$.

\noindent
The separation
\begin{eqnarray}
\sigma^{\rm virt} + \sigma^{\rm soft}   = 
\tilde\sigma + \sigma_{\rm remainder} \, , 
\label{eqn:QED:1}
\end{eqnarray}
\begin{eqnarray}
\tilde\sigma  =
\frac{\alpha}{\pi}\,
\left[
(L_e-1)
\log\frac{4 \Delta E^2}{s} +
\frac{3}{2}\,L_e
\right]  \cdot \sigma^{\rm Born}\, ,  \nonumber
\end{eqnarray}
identifies a one-loop contribution $\sigma_{\rm remainder}$ 
that is IR finite and free of large universal QED terms.
$\sigma_{\rm remainder}$ contains 
the MSSM-specific radiative corrections,
whereas the subtracted part $\tilde\sigma$ in~(\ref{eqn:QED:1})
together with the hard bremsstrahlung part 
from initial-state radiation can be considered
as a contribution of the type ``QED corrections'',
\begin{eqnarray}
\sigma_{\rm QED} & = &
\sigma^{\rm hard}_{\rm ISR} + \tilde\sigma \, .
\label{eqn:QED:3}
\end{eqnarray}
It is independent of the auxiliary cut $\Delta E$, and it
contains all large logarithms from virtual, soft, and hard photons.

With this reordering,
the complete cross section at the one-loop level
can be written  as follows,
\begin{eqnarray}
\sigma^{\rm 1-loop} & = &
\sigma^{\rm Born} + \sigma^{\rm virt} + 
\sigma^{\rm soft} + \sigma^{\rm hard} \\ & = &
\sigma^{\rm Born} + \sigma_{\rm QED} + \sigma_{\rm remainder}
+\Delta\sigma . \nonumber
\end{eqnarray}
The small piece $\Delta\sigma$ accounts for final-state radiation
in case of chargino production, and for photon radiation from the $t$-channel
selectrons in case of neutralino production.

\subsection{Results}

\begin{figure}
\begin{center}
\psfrag{Process}{$e^+e^- \to \tilde\chi^+_1 \, \tilde\chi^-_1 \, (\gamma)$}
\psfrag{Spb}{$\sigma/{\rm pb}$}
\psfrag{EGeV}{$\sqrt{s}/{\rm GeV}$}
\psfrag{Born}{$\sigma^{\rm Born}$}
\psfrag{Remainder}{$\sigma_{\rm remainder}$}
\psfrag{QED}{$\sigma_{\rm QED}$}
\psfrag{Complete}{$\sigma^{\rm 1-loop}$}
\rotatebox{-90}{
\scalebox{0.50}[0.56]{
  \epsfbox{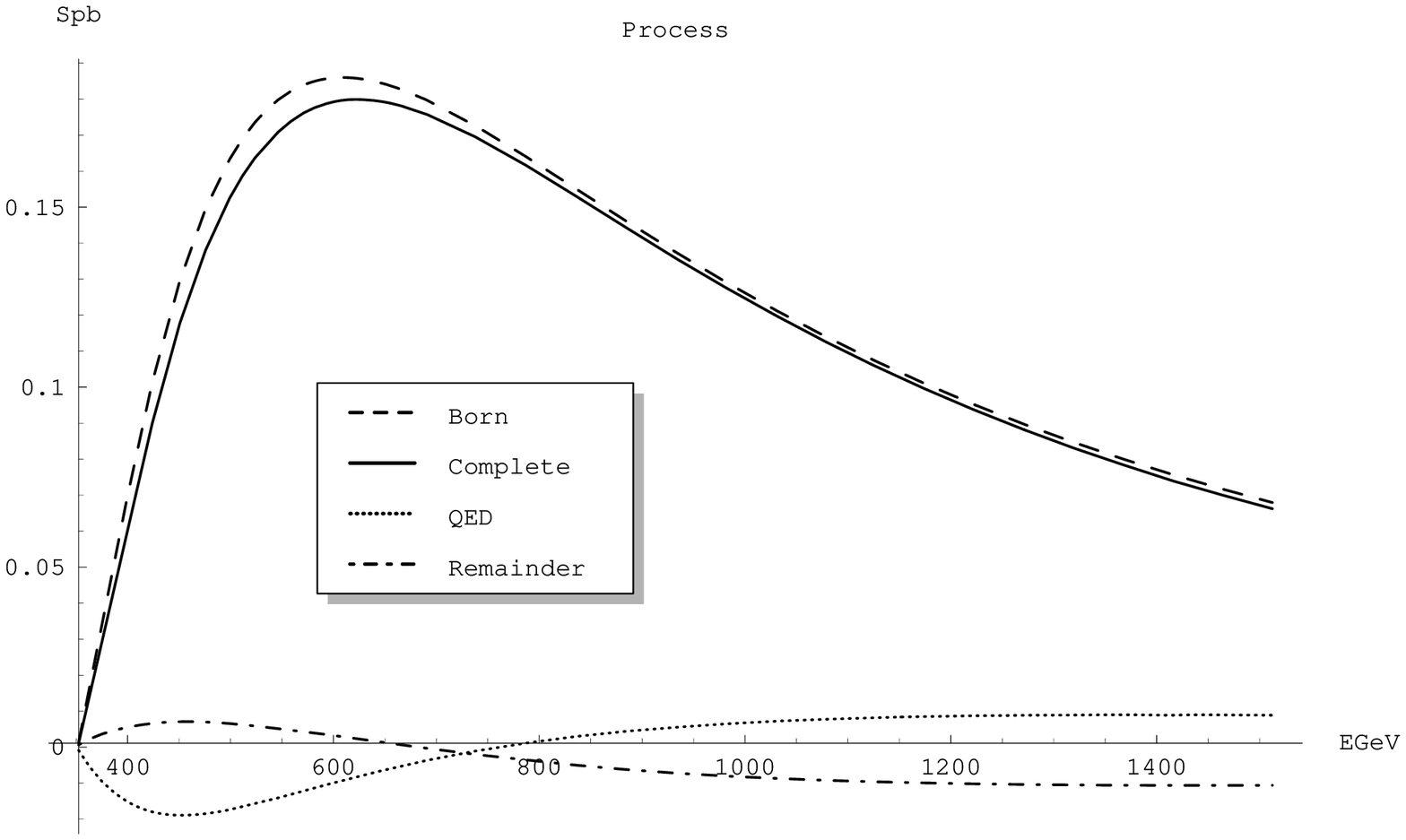}
  \psfrag{Process}{$e^+e^- \to \ntrl{1} \, \ntrl{2} \, (\gamma)$}
  \epsfbox{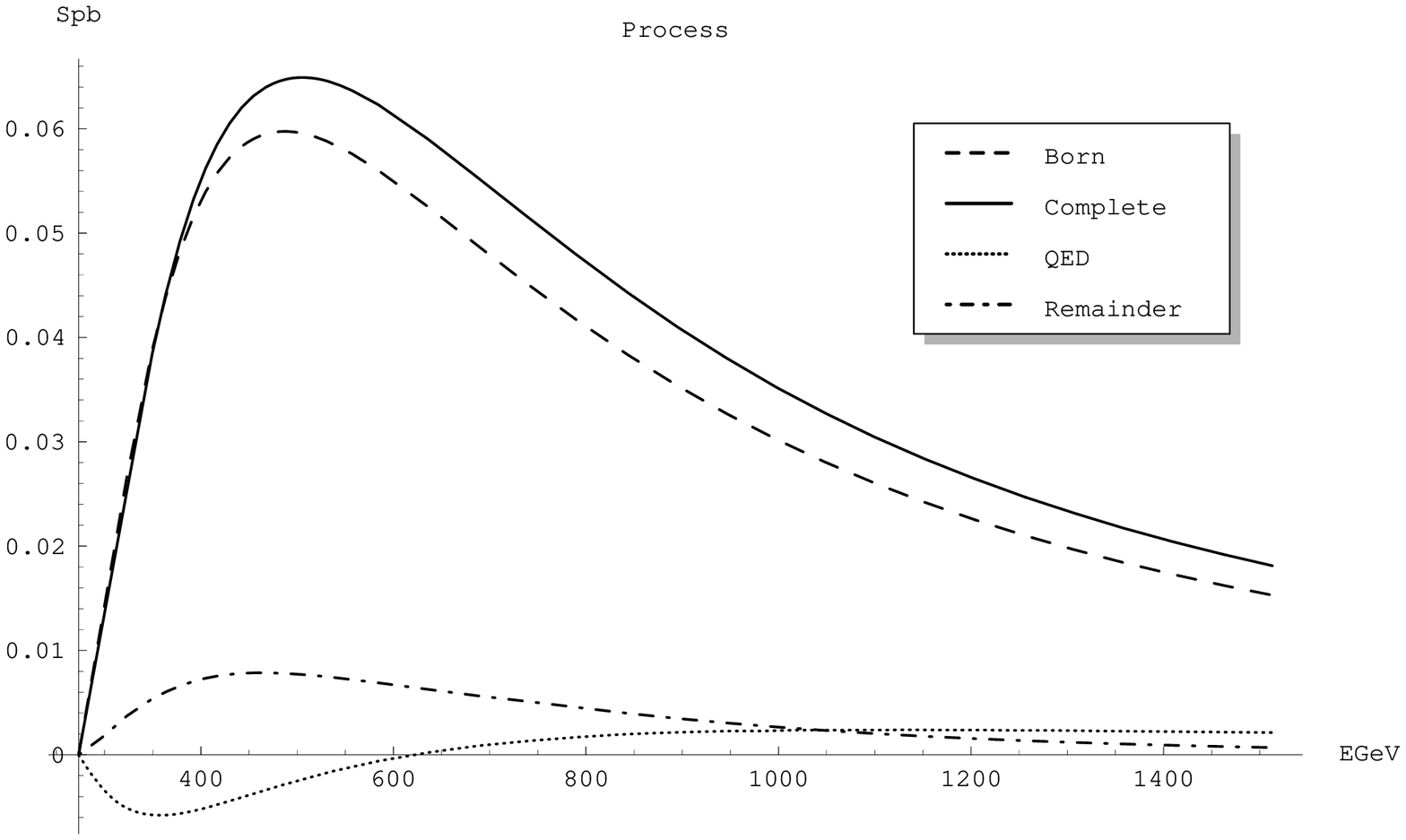}
}
}

\end{center}
\end{figure}

For illustration, we choose the example of the 
SPS1a parameter set~\cite{sps1a} listed partially in section~3.
The $\overline{\rm DR}$ parameters are first converted into 
on-shell quantities and pole masses, from which 
the cross sections are calculated, both at tree level and 
at higher order.
The following figures contain the integrated cross sections
for $\tilde\chi^+_1 \tilde\chi^-_1$ and for
$\tilde\chi^0_1 \tilde\chi^0_2$ 
production in $e^+e^-$ annihilation
for unpolarized beams,
as functions of the CMS energy. Besides Born cross sections and full
one-loop results, also the contributions $\sigma_{\rm remainder}$
and $\sigma_{\rm QED}$ are shown separately.

\end{document}